\documentclass[11pt]{article}
\usepackage{dcolumn}
\usepackage{graphicx}
\usepackage{lineno}
\newcommand{\etal}{{\it et al.}}

\newcommand{\dspm}{\ensuremath{D^\pm_s}}
\newcommand{\dpm}{\ensuremath{D^\pm}}
\newcommand{\pipipipm}{\ensuremath{\pi^{-}\pi^{+}\pi^{\pm}}}
\newcommand{\kkpipm}{\ensuremath{K^{-}K^{+}\pi^{\pm}}}

\newcommand{\Dkpipi}{\ensuremath{D^{+}\to K^{-}\pi^{+}\pi^{+}}}
\newcommand{\Dkkpi}{\ensuremath{D^\pm\to K^{+}K^{-}\pi^\pm}}

\newcommand{\Dskkpi}{\ensuremath{D^{+}_s\to K^{+}K^{-}\pi^{+}}}
\newcommand{\Dspipipi}{\ensuremath{D_s^{\pm}\to \pi^{+}\pi^{-}\pi^{\pm}}}

\newcommand{\CP}{\ensuremath{CP}}

\newcommand{\gevcccc}{\ensuremath{{\mathrm{\,Ge\kern -0.1em V^2\!/}c^4}}}

\title{Amplitude Analyses of D Decay Dalitz Plots}
\author{Milind V. Purohit\\ Univ. of S. Carolina\\ representing the BaBar collaboration}
\date{September 30, 2012}

\begin{document}
\maketitle

\begin{abstract}
We describe Dalitz plots arising in studies of $D$-meson decays, and what
we have learned from them in recent years. After some mention of
technique and history, we focus on decay of the \dpm\ and \dspm\
mesons to the \kkpipm\ final state, and on \dspm\ decay to
\pipipipm. Some mention of other Dalitz plots involving $D$-meson decays
is also made.
\end{abstract}

\newpage

\section{Introduction: General observations on Dalitz plots}

\begin{figure}[tbp]
  \centerline{\scalebox{1.5}{\includegraphics{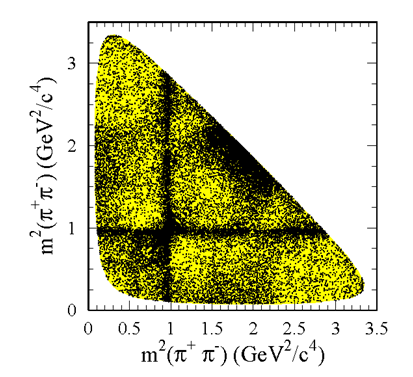}}}
  \caption{BaBar data~\cite{babar_ds_3pi}: Dalitz plot for \Dspipipi\ decays.}\label{fig:babar_ds_3pi}
\end{figure}

\begin{figure}[tbp]
  \centerline{\scalebox{1.5}{\includegraphics{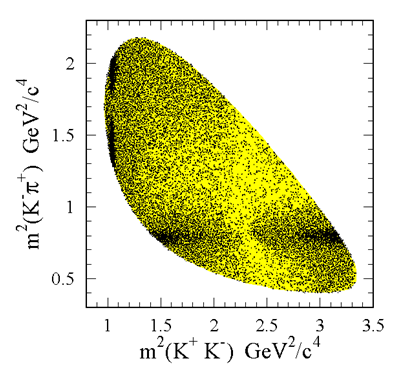}}}
  \caption{BaBar data~\cite{babar_ds_kkpi}: Dalitz plot for \Dskkpi\ decays.}\label{fig:babar_ds_kkpi}
\end{figure}

Typically, Dalitz plots are scatterplots of two kinematic variables
describing the decay of a pseudo-scalar (P) meson to three
pseudo-scalars, i.e., $P\to PPP$ decays.
Figures~\ref{fig:babar_ds_3pi}, \ref{fig:babar_ds_kkpi},
\ref{fig:babar_kkpi}, and \ref{fig:lhcb_kkpi} show examples of such
plots. Such kinematics should be described by 9 variables (for the three
final state 3-momenta), but energy-momentum conservation cuts this down
to five variables. Further, the lack of spin in the parent and daughters
leads to a lack of preferred direction in space; thus two angles
describing the direction of emission of any one daughter in the parent
rest frame (RF) are inconsequential as is the angle described by the
decay plane when rotated with this direction as the axis. Thus we reduce
the five variables down to two non-trivial variables. The decay phase
space can be shown to be uniform in either the energies of any two of
the daughters in the parent RF, or in the mass-squareds of any two
daughter pairs. Of course, in each case, the third corresponding
variable is kinematically related to the two that are chosen for display
in the plot. Finally, we should note that in case two or more daughters
are identical the decay amplitude must be symmetric under interchange of
those two, which leads to a symmetry in the Dalitz plot which can either
be explicitly seen or the plot can be folded.

As we will see below the predicted uniform phase space of Dalitz plots
can be exploited to bring out the dynamical features of the decays. For
instance, resonances in the decay chain stand out as peaks in the
two-dimensional plot (and are particularly easy to identify and study
for the mass-squared version). Thus, Dalitz plot studies can be used to
find resonances and to measure their parameters such as their masses and
widths. Finally, since Dalitz plots reveal interference of decay
amplitudes in such a spectacular way they may be used to search for
CP-violating effects which also arise due to interference of decay
amplitudes. 

\section{Fits to Dalitz plots}
The decay amplitude squared leads to the distribution visible in a
Dalitz plot. Parameters for such an amplitude can be fit using an
using either a binned or, preferably, an unbinned fit. 

A typical fit utilizes a model which describes the amplitude as a sum
over complex amplitudes as follows:

\begin{equation}
  {\cal A} = {\cal A}_0 + \sum_{\ell = 1} {\cal A}_{\ell i} 
\label{eq:iso1}
\end{equation}

where ${\cal A}_0$ is the $S$-wave amplitude and each ${\cal A}_{\ell
  i}$ is the amplitude describing the $i^{th}$ resonance of spin
$\ell$. Such a model is often called the ``isobar'' model for historical
reasons, and the individual resonant amplitudes ${\cal A}_{\ell i}$ are
in turn described by expressions such as

\begin{equation}
  {\cal A}_{\ell i} = \alpha_{\ell i} \; \exp{(i\beta_{\ell i})} \; 
                      {\cal D} \; {\cal R}_{\ell i} \; 
                      {BW_{\ell i}} \; {\cal Q}_{\ell i}
\label{eq:iso2}
\end{equation}

where $D$ and ${\cal R}_{\ell i}$ are form factors of the $D$-meson and
the resonance $R$, and where $BW_{\ell i}$ and ${\cal Q}_{\ell i}$ are
the resonance Breit-Wigner (propagator) and spin factors,
respectively. Of course, each of these factors is a function of
kinematic quantities which are suppressed here in the interest of brevity.

The fits to data require that we integrate the differential decay rate
over the phase space, and a simplyfying factor is that most of the
integrals can be done just once since the complex coefficients
$\alpha_{\ell i}\exp{(i\beta_{\ell i})}$ factor out. Resonance
parameters sometimes cause problems in the fitting procedure because
integrals then have to be done every time these are changed.

Background contributions to the Dalitz plot have to be estimated either
from data or Monte Carlo (MC). In principle, the $D$-meson mass
sidebands in data are a better source of knowledge for backgrounds, but
do not fit kinematically into exactly the same Dalitz plot. If we apply
a $D$-mass constraint to these background regions to solve the kinematic
problem with the plot boundaries, the two-body masses are slightly
shifted, which causes problems in describing resonances. All these
effects must be properly accounted for before a successful fit can be
done. Variations in detector efficiency across the Dalitz plot must also
be modeled in some way, typically as a smooth function of position in
the plot. This is often done using a polynomial or an Artificial Neural
Net fit to the data.

Finally, we should elaborate on the $S$-wave term ${\cal A}_0$. While
this is often taken to be just a complex constant, more recently
experiments such as E791~\cite{e791_kpipi_06} have experimented with a
mass-dependent piece-wise linear ``function'', or set of constants,
instead. Surprisingly, this works reasonably well in the fits, i.e., the
resulting constants vary smoothly with mass. This has come to be known
as the ``Model Independent Partial Wave Analysis'' (MIPWA) technique.

\section{Early $D$ Dalitz results}

Since there are three flavors of $D$-meson ($D^0$, $D^+$ and $D^+_s$),
and many possible daughter particles (neutral and charged $\pi$, $K$) a
large number of Dalitz plots can arise in $D$-decays, but only a few of
these have been studied. Such plots were first made in the 1980's and
1990's by the Mark II~\cite{mark2_kpipi} experiment which observed a
``non-uniform'' density, by the Mark III~\cite{mark3_kpipi} experiment
which published a Dalitz plot analysis, and by the E691~\cite{e691_93}
experiment which confirmed the main features of the previous fit to
\Dkpipi\ decays viz., the presence of a strong non-resonant amplitude, 
and the fact that the isobar model gives a poor overall fit.

The fit problems were generally traced to the $S$-wave amplitudes and
explored further by the E791 experiment. Early work in E791 resolved the
fit issues by introducing a $\kappa(800)$ resonance for the $K^-\pi^+$
system~\cite{e791_kpipi_02}, and a $\sigma(500)$
resonance~\cite{e791_3pi} for the $\pi^-\pi^+$ system.

Later, following a suggestion by W.M. Dunwoodie of SLAC, E791 pursued a
model-independent way to compare Dalitz decay data to LASS results on
$K^-\pi^+$ scattering~\cite{lass_88}. This E791 study, pioneered by
B.~Meadows, proved to be a success for this MIPWA
technique~\cite{e791_kpipi_06}. In the meanwhile, the FOCUS experiment
found an acceptable fit to \Dkpipi\ decays using a K-matrix description
of the S-wave with no $\sigma(500)$ pole.

\section{Recent $D$ Dalitz Plot results}

More recently, higher statistics data became available from the CLEO,
CLEO-c, BaBar and Belle experiments. Some of these fits are described
below. Also, $D$-decay data from these experiments have been
used to study CP-violation in $D$ decays and also in $B$ decays.

\begin{figure}[tbp]
  \centerline{\scalebox{1.5}{\includegraphics{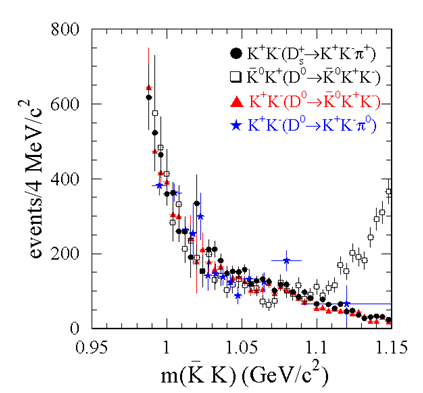}}}
  \caption{BaBar data: $K\overline{K}$ S-wave intensity in Dalitz decays
    as a function of $m_{KK}$~\cite{babar_ds_kkpi}.}\label{fig:kk_amp}
\end{figure}

Shown in Fig.~\ref{fig:babar_ds_kkpi} is the \Dskkpi\ Dalitz plot from
BaBar~\cite{babar_ds_kkpi}. Using moments of Legendre polynomials of
helicity angles, BaBar extracted the $S$, $P$ magnitudes and relative
phase in the region (0.99 $<$ $m_{K^-K^+}$ $<$ 1.15) GeV/$c^2$ and found
that the $P$-wave is essentially pure $\phi(1020)$, and is therefore
described as such to extract a binned S-wave amplitude. Comparing
$S$-wave intensities from different Dalitz plots (see
Fig.~\ref{fig:kk_amp}) one finds good agreement except for the
$\overline{K}^0K^+$ amplitude above 1.1
GeV/$c^2$. Figure~\ref{fig:babar_ds_3pi} shows the BaBar \Dspipipi\ decays
Dalitz plot~\cite{babar_ds_3pi}.

\begin{figure}[tbp]
  \centerline{\scalebox{0.10}{\includegraphics{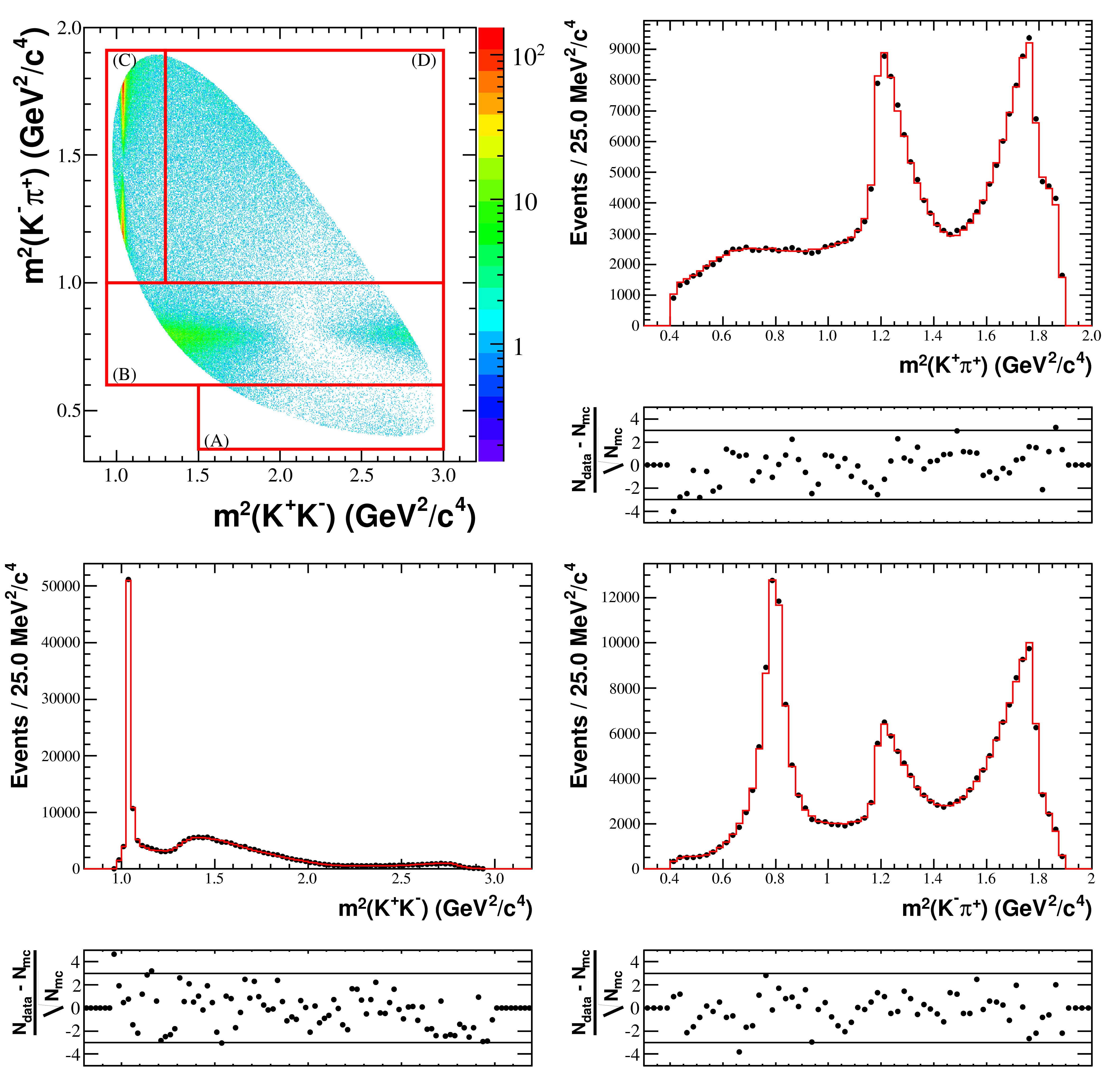}}}
  \caption{BaBar preliminary data: \Dkkpi\ Dalitz plot fit projections
    assuming no $CP$V, with the regions used for model-independent
    comparisons indicated by boxes on the scatterplot. The data are
    represented by points with error bars, the fit results by the
    histogram. The normalized residuals shown below, defined as
    $(N_{Data} - N_{MC})/\sqrt{N_{MC}}$, lie within a $\pm 5\sigma$ band
    in each case, with almost all of the values within $\pm
    3\sigma$.}\label{fig:babar_kkpi}
\end{figure}

\begin{figure}[tbp]
  \centerline{\scalebox{1.2}{\includegraphics{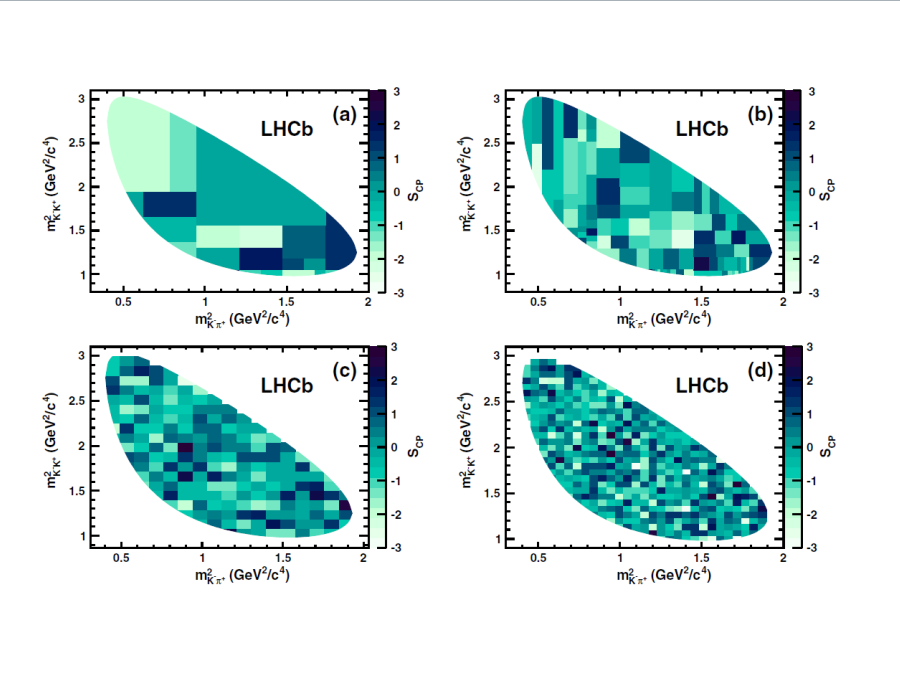}}}
  \caption{LHCb data: CP\ asymmetry in progressively smaller
    bins of the Dalitz plot for \Dkkpi\ decays.}\label{fig:lhcb_kkpi}
\end{figure}

\section{Search for CPV in \Dkkpi\ decays}

Most recently there has been an analysis from BaBar on the search for
CP-violation in DCS \Dkkpi\ decays.  Figure~\ref{fig:babar_kkpi} shows
the combined Dalitz plot for these decays.  Searches for $CP$ violation
($CP$V) in charm meson decays provide a probe of physics beyond the
Standard Model (SM).  Singly Cabibbo-suppressed (SCS) decays can exhibit
direct \CP\ asymmetries due to interference between tree-level
transitions and $\Delta C = 1$ penguin-level transitions if there is
both a strong and weak phase difference between the two amplitudes. In
the SM, the resulting asymmetries are suppressed by
$\mathcal{O}(|V_{cb}V_{ub}/V_{cs}V_{us}|)\sim10^{-3}$.  Measurement of a
larger \CP\ asymmetry could be caused by the enhancement of penguin
amplitudes in $D$ decays due to final-state
interactions~\cite{GronauRosner,ChengChiang} or by New
Physics~\cite{GKN07,AMP08}.

The LHCb and CDF collaborations recently reported evidence for the
difference in \CP\ asymmetry of the time-integrated $D^0\to \pi^+\pi^-$
and $D^0\to K^+K^-$ decay rates~\cite{LHCb, CDF}. In the decay mode
\Dkkpi\ the CLEO-c collaboration has reported~\cite{cleoc_kkpi} that the
asymmetry defined by 
\begin{equation}
  {\cal A}_{CP} = \int dm^2_+dm^2_-\,{
                      |{\cal M}|^2 - |\overline{\cal M}|^2 
                                    \over 
                      |{\cal M}|^2 + |\overline{\cal M}|^2 
                                   } \quad
               / \quad \int dm^2_+dm^2_-
\label{eq:cleoc_asym}
\end{equation}
is found to be $(-0.4\pm 2.0^{+0.2\,+0.6}_{-0.5\,-0.3})\%$ i.e.,
consistent with zero.  The LHCb collaboration has also reported on a
search in this decay mode; Fig.~\ref{fig:lhcb_kkpi} shows the asymmetry
in various-sized regions of the Dalitz plot from that
experiment~\cite{lhcb_kkpi}. After taking account of the overall rate, 
they do not find any evidence for \CP\ violation in the differential
decay rate. Searches for \CP\ violation in other SCS decays with
identical transitions $c\to ud\bar{d}$ and $c\to u s \bar{s}$ are
relevant for understanding the origin of {\CP}V~\cite{GKZ12, GIP12,
  FMS12}.

\begin{figure*}[!tb]
\begin{center}
\includegraphics[width=0.3\textwidth]{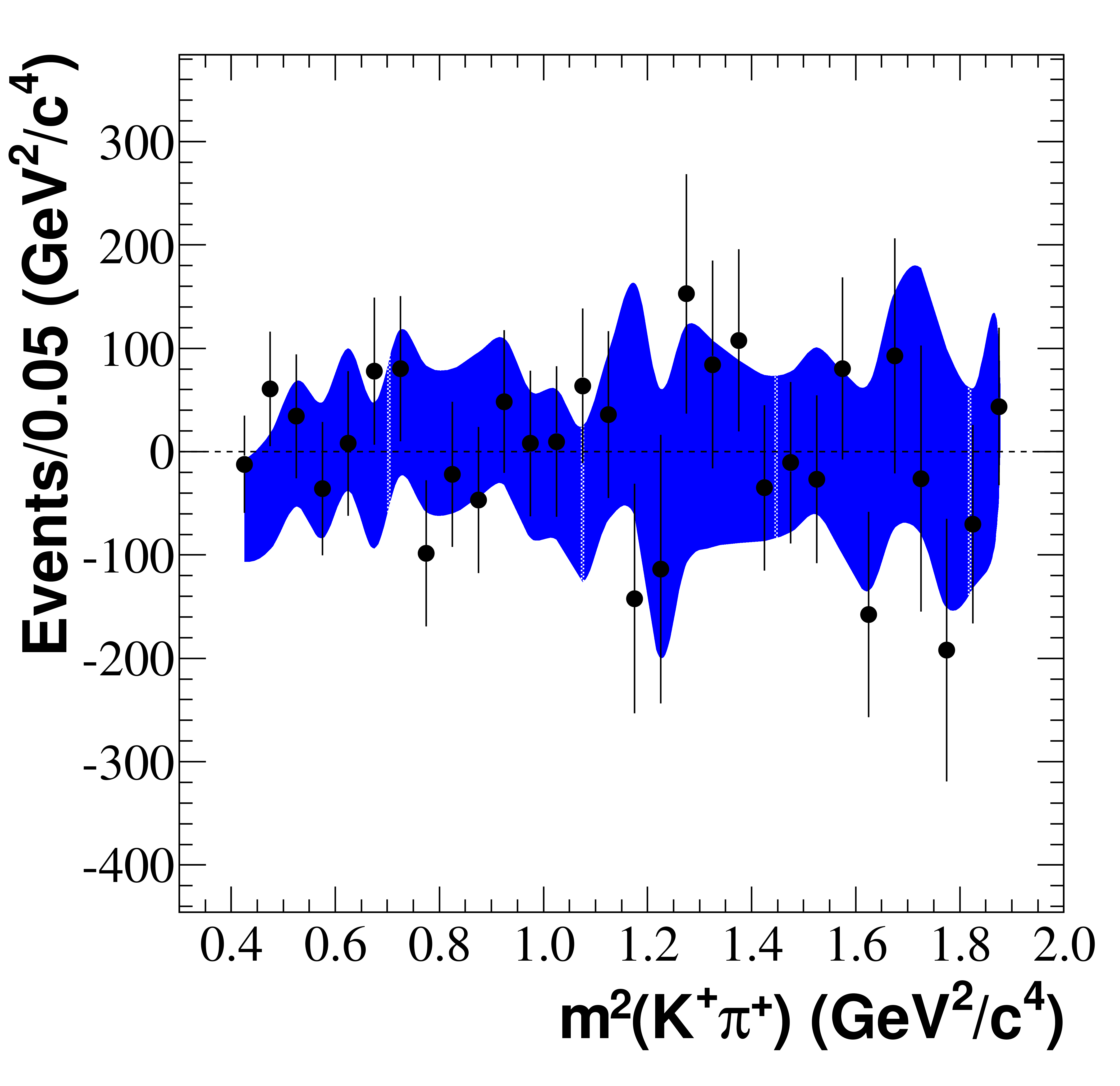}
\includegraphics[width=0.3\textwidth]{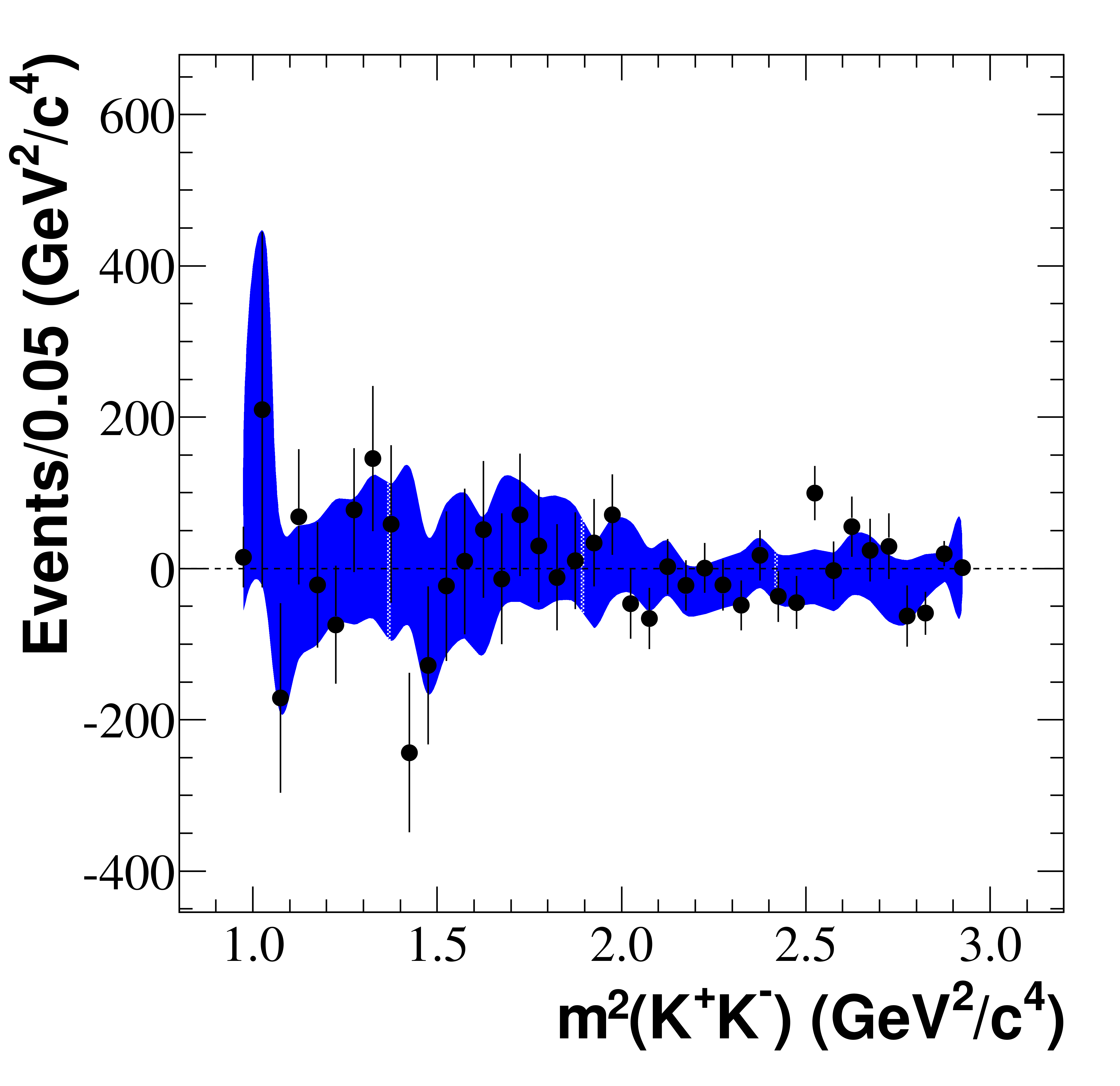}
\includegraphics[width=0.3\textwidth]{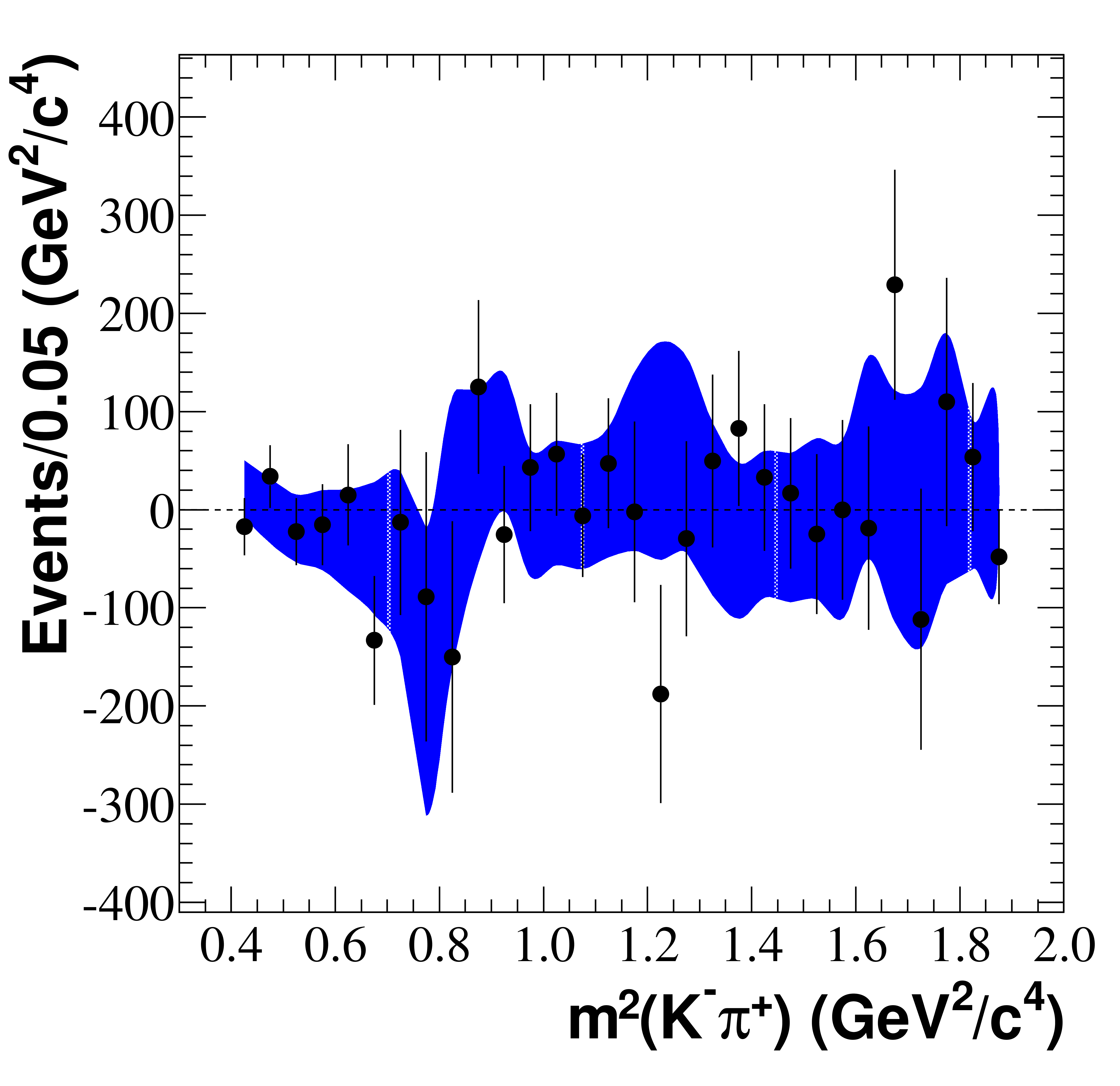}
\vspace{-0.3cm}
\caption{BaBar preliminary data: The difference of the Dalitz plot
    projections of data (points) and the fit (blue band) between the
    $D^+$ and $D^-$ decays. The width of the band represents the $\pm
    1\sigma$ error expected for our data sample size.}\label{fig:kkpi_cp_vs_msq}
\vspace{-0.7cm}
\end{center}
\end{figure*}

The BaBar analysis does not find any evidence for \CP\ violation in the
SCS decay \Dkkpi. The integrated \CP\ asymmetry is found to be
$(0.35\pm0.30\pm0.15)\%$. We find also that the asymmetries in four
regions of the Dalitz plot are consistent with zero, and that the $D^+$
and $D^-$ Dalitz plots are consistent with no \CP\ asymmetry with a
probability of 72$\%$, according to the analysis of the normalized
residuals of the $D^+$ and $D^-$ Dalitz plot divided into 100 equally
populated bins. Finally, we find no evidence for \CP\ asymmetry in
decays through various intermediate states with a study of the two-body
mass distributions as seen in Fig.~\ref{fig:kkpi_cp_vs_msq}, and with a
parameterization of the Dalitz plot for which the measured
\CP\ asymmetries in amplitudes are listed in Table~\ref{tab:dpmodel}.

\begin{table*}
\centering
\setlength{\extrarowheight}{5pt}
\caption{Preliminary \CP\ violating parameters from the Dalitz plot fit. The first
  errors are statistical, the second are systematic uncertainties which are
  determined from taking in quadrature the errors associated with
  tracking, the production model correction, the event selection, and
  the Dalitz plot model.}
\begin{tabular}{lccc}
\hline 
\hline
Resonance & r ($\%$) & $\Delta\phi\;(^{\circ})$\\ 
\hline
$\bar{K}^{*}(892)^{0}$          & 0. (FIXED)                                    & 0. (FIXED)                            \\
$\bar{K}^{*}_{0}(1430)^{0}$     & $-9.40^{+5.65}_{-5.36} \pm 4.42$   & $-6.11^{+3.29}_{-3.24} \pm 1.39$   \\
$\phi(1020)$                    & $0.35^{+0.82}_{-0.82} \pm 0.60$    & $7.43^{+3.55}_{-3.50} \pm 2.35$    \\ 
NR                              & $-14.30^{+11.67}_{-12.57} \pm 5.98$ & $-2.56^{+7.01}_{-6.17} \pm 8.91$  \\
$\kappa(800)$                   & $2.00^{+5.09}_{-4.96} \pm 1.85$    & $2.10^{+2.42}_{-2.45} \pm 1.01$    \\ 
$a_{0}(1450)^{0}$               & $5.07^{+6.86}_{-6.54} \pm 9.39$            & $4.00^{+4.04}_{-3.96} \pm 3.83$ \\
\hline
		& $\Delta x$                            & $\Delta y$                    \\              
$f_{0}(980)$                    & $-0.199^{+0.106}_{-0.110} \pm 0.084$ & $-0.231^{+0.100}_{-0.105} \pm 0.079$ \\
$f_{0}(1370)$                   & $0.019^{+0.049}_{-0.048} \pm 0.022$        & $-0.0045^{+0.037}_{-0.039} \pm 0.016$ \\

		    \hline
\hline
\end{tabular} 
\label{tab:dpmodel}
\end{table*}

\section{Acknowledgements}
I would like to thank my collaborators on BaBar and at SLAC for making
this talk possible, and particularly Brian Meadows and Bill Dunwoodie
for all their innovative work on Dalitz plots. Most of all though, I
would like to thank Ryan White whose tireless work on the search for CP
violation in \Dkkpi\ decays has set a new bar for this kind of
study. This work was made possible by a grant from the U.S. Department
of Energy.

\end{document}